\documentclass[%
 reprint,
superscriptaddress,
 amsmath,amssymb,
 aps,
]{revtex4-2}

\usepackage{xcolor}
\usepackage{graphicx}% Include figure files
\usepackage{dcolumn}% Align table columns on decimal point
\usepackage{bm}% bold math
\usepackage{hyperref}% add hypertext capabilities
\renewcommand{\selectlanguage}[1]{}

\begin{document}

\preprint{APS/123-QED}

\title{Exploiting separation-dependent coherence to boost optical resolution}

\author{Ilya Karuseichyk}
\email{ilya.karuseichyk@lkb.upmc.fr}
\affiliation{Laboratoire Kastler Brossel, Sorbonne Universit\'{e}, CNRS, ENS-Universit\'{e} PSL, Coll\`{e}ge de France, 4 place Jussieu, F-75252 Paris, France
}
\author{Giacomo Sorelli}
\affiliation{Fraunhofer IOSB, Ettlingen, Fraunhofer Institute of Optronics,
System Technologies and Image Exploitation, Gutleuthausstr. 1, 76275 Ettlingen, Germany}
\author{Vyacheslav Shatokhin}
\affiliation{Physikalisches Institut, Albert-Ludwigs-Universität Freiburg, Hermann-Herder-Straße 3, D-79104 Freiburg, Germany}
\affiliation{EUCOR Centre for Quantum Science and Quantum Computing, Albert-Ludwigs-Universität Freiburg, Hermann-Herder-Straße 3, D-79104 Freiburg, Germany}
\author{Mattia Walschaers}
\affiliation{Laboratoire Kastler Brossel, Sorbonne Universit\'{e}, CNRS, ENS-Universit\'{e} PSL, Coll\`{e}ge de France, 4 place Jussieu, F-75252 Paris, France
}
\author{Nicolas Treps}
\affiliation{Laboratoire Kastler Brossel, Sorbonne Universit\'{e}, CNRS, ENS-Universit\'{e} PSL, Coll\`{e}ge de France, 4 place Jussieu, F-75252 Paris, France
}

\date{\today}% It is always \today, today,
             %  but any date may be explicitly specified

\begin{abstract}
    The problem of resolving point-like light sources not only serves as a benchmark for optical resolution but also holds various practical applications ranging from microscopy to astronomy. In this research,  we aim to resolve two thermal sources sharing arbitrary mutual coherence using the spatial mode demultiplexing technique. Our analytical study includes scenarios where the coherence and the emission rate depend on the separation between the sources, and is not limited to the faint sources limit.  
    We consider the fluorescence of two interacting dipoles to demonstrate that the dependence of emission characteristics on the parameter of interest can boost the sensitivity of the estimation and noticeably prolong the duration of information decay.
\end{abstract}
 
 \maketitle

\section{Introduction}
The role of coherence in imaging problems has been intensively studied and discussed for decades \cite{considine_effects_1966,nayyar_two-point_1978,aert_resolution_2006}. This topic became especially interesting with the formulation of the imaging problem in terms of parameters estimation \cite{helstrom_image_1967,helstrom_resolvability_1970}. A widely studied form of this problem is the estimation of the separation between two point sources of light, which not only serves as a benchmark for optical resolution but also has numerous practical applications ranging from microscopy to astronomy \cite{baddeley_biological_2018,tsang_resolving_2019}. The parameter estimation approach to this simple imaging problem allows to find an ultimate quantum limit for the resolution \cite{tsang_quantum_2016,nair_far-field_2016,lupo_ultimate_2016,sorelli_quantum_2022}. The Fisher information of the spatial-mode demultiplexing (SPADE) technique was shown to saturate the quantum limit for incoherent \cite{tsang_quantum_2016,nair_far-field_2016, rehacek_optimal_2017} and partially coherent \cite{tsang_resurgence_2019} faint sources. The sensitivity measure based on the method of moments \cite{gessner_metrological_2019} allowed to study a more general case of bright incoherent  \cite{sorelli_moment-based_2021,sorelli_optimal_2021}, fully coherent \cite{karuseichyk_resolving_2022} and entangled \cite{zhang_quantum_2023} sources, demonstrating non-trivial scaling of the sensitivity with the brightness of the thermal sources and showing that the SPADE approach remains quantum optimal in this more general scenario. Moreover, the method of moments provides a simple estimation strategy for the parameter of interest that relies solely on the measured mean intensities and does not require maximum likelihood estimation or other heavy computations. At the same time, the sensitivity obtained from the method of moments saturates the Cramer-Rao bound in the limit of faint sources. 

In this study we analyze the case of partially coherent sources that sparked hot debates in the community \cite{larson_resurgence_2018,tsang_resurgence_2019,larson_resurgence_2019,hradil_quantum_2019,hradil_exploring_2021,liang_coherence_2021,liang_quantum_2023,kurdzialek_back_2022,wang_quantum-limited_2023}. We aim to resolve two bright thermal sources sharing arbitrary mutual coherence, including cases when the latter depends on the separation between the sources. This situation was partially covered in Ref.\cite{wang_quantum-limited_2023} using the approximation of faint sources and assuming a lossless optical system. In this contribution, we instead consider sources with arbitrary brightness, since the approximation of faint sources is not always applicable in practical situations. 
Furthermore, previous findings indicate that the lossless model tends to underestimate sensitivity in cases where the initial brightness of the sources is known  \cite{tsang_resurgence_2019,kurdzialek_back_2022}. To address this limitation, our model explicitly incorporates losses in the imaging system. 

Our approach, employing the method of moments, extends to scenarios where both mutual coherence and the emission rate of the sources are dependent on the separation. This method allows us to address a practically significant issue of resolving two reflectors illuminated by a source of light with finite coherence width, where a faint source limit is not always applicable. We show that the finite coherence width of the illumination can significantly enhance the optical resolution, however, this enhancement becomes less pronounced with the growing number of coherently emitted photons, i.e. for a more extended coherence time of the illumination source. 

 In the second example, we resolve the separation between two interacting dipole emitters. We show that the dependence of emission characteristics on the separation can drastically boost the estimation sensitivity. As a result, one can obtain useful information about the dipoles' separation on a timescale significantly larger than the radiative lifetime of the excited state of the dipole. For the late stages of decay, even though the probability of detecting a photon is very small, the detection event contains a lot of information because of the strong interaction-induced correlations between the emitters. {We show that this effect is resistant to the dephasing of the emitters and to the detection noise.}%Consequently, due to the interaction between emitters sensitivity of the separation estimation can increase for several orders of magnitude.

\section{The model of emitters}
\begin{figure}
    \centering
    \includegraphics[width=.75\linewidth]{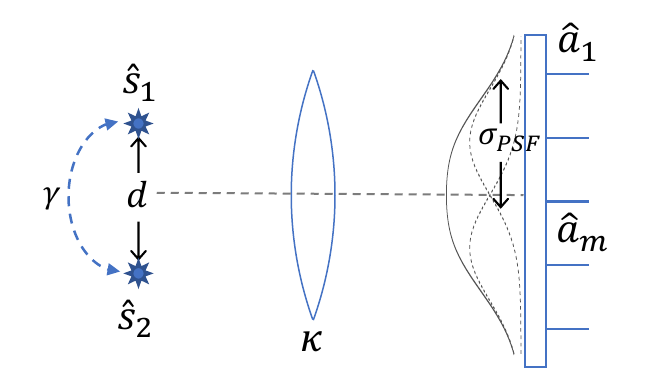}
    \caption{The scheme of the two point sources resolving. The sources emit light into orthogonal modes $\hat s_{1,2}$ with a mutual coherence  $\gamma$. The light propagates through the lossy imaging system with transmissivity $\kappa$ and PSF width $\sigma_{PSF}$ before being decomposed into spatial modes $f_m(x)$ with corresponding field operators $\hat a_m$ and subsequently detected. }
    \label{fig:scheme}
\end{figure}

We consider a traditional optical scheme for single-parameter imaging in which the only estimated parameter is the separation $d$ between two point-like light sources (Fig. \ref{fig:scheme}). All other parameters of the sources are assumed to be known, i.e. the imaging setup can be aligned in such a way that both sources lie on its X-axis and the centroid of the sources lies at the coordinate origin. Thus the problem can be treated in one dimension with coordinates of the sources being $x_j=(-1)^j d/2$. The sources emit light into orthogonal modes with field operators $\hat s_{1,2}$. Each of these modes describes a spherical wave emitted by the corresponding source during its coherence time.  We only consider equally bright sources emitting thermal light, including cases of correlated emission. Since these stochastic sources do not have any external phase reference
, the emitted state should be symmetric over the global phase shift. This implies that  $\langle \hat s_j \rangle = 0$ and $\langle \hat s_j \hat s_k \rangle = 0$. Thus the state is fully characterized by the first-order coherency matrix 
\begin{equation}
    \label{eqn:cohS}
\Gamma^{(1)\hat s}_{j k}=\langle \hat s^\dagger_j \hat s_k \rangle=
    \begin{pmatrix}
    N & \gamma N\\
     \gamma^* N & N
        \end{pmatrix},
\end{equation}
where $N$ is the average number of photons emitted by each source per its coherence time. Assuming the coherence time of the sources to be fixed we refer to $N$ as brightness and define sources with $N \ll 1$ as faint sources. Note, however, that the number of photons emitted by such sources per integration time of the detection system can still be large if the coherence time of the sources is short compared to the integration time.  Unlike most research on this topic, our approach allows for studying bright thermal sources with $N>1$, demonstrating the non-trivial scaling of the separation estimation sensitivity with this parameter. The complex number $\gamma=\gamma_0 e^{i \phi}$ with $-1 \le \gamma_0 \le 1$ and $0 \le \phi < \pi$ stands for the mutual coherence between the sources. In the general case, $\gamma$ can depend on the separation. The presence of a non-diagonal part in the coherency matrix corresponds to correlations of the quadratures
\begin{equation}
\begin{split}
\label{eqn:V}
   V_{jk} =&\frac{1}{2} \left\langle \{\hat Q_j, \hat Q_k\} 
   %+ \hat Q_k \hat Q_j 
   \right\rangle = \\ 
   %-   \left\langle \hat Q_j  \right\rangle  \left\langle  \hat Q_k   \right\rangle
   =&
    \begin{pmatrix}
    N+\frac{1}{2}& 0 & N \operatorname{Re} \gamma  & N \operatorname{Im} \gamma \\
    0 & N+\frac{1}{2} & -N \operatorname{Im} \gamma & N \operatorname{Re} \gamma \\
     N \operatorname{Re} \gamma  & -N \operatorname{Im} \gamma &
     N+\frac{1}{2} & 0 \\
    N \operatorname{Im} \gamma & N \operatorname{Re} \gamma &  
    0 & N+\frac{1}{2}
    \end{pmatrix},
\end{split}
\end{equation}
with quadrature vector $\hat{\Vec{{Q}}}=(\hat q_1,\hat p_1,\hat q_2,\hat p_2)^T,$ with $\hat q_j=( \hat s^\dagger_j + \hat s_j )/\sqrt{2}$ and $\hat p_j=i( \hat s^\dagger_j - \hat s_j )/\sqrt{2}$.
%and absence of the mean-field $\langle \hat Q_j \rangle=0$ is taken into account.
In this case mutual coherence originates from thermal correlations, as opposed to coherent states where mutual coherence comes from a nonzero mean-field. 

\section{Field transformation through the imaging system}

Light emitted by the point sources propagates through a diffraction-limited imaging system with transmissivity $\kappa$ (see Fig. \ref{fig:scheme}). The corresponding transformation of the field operators in the paraxial approximation can be represented  as~\cite{lupo_ultimate_2016}
\begin{equation}
    \hat c_j = \sqrt \kappa \hat s_j + \sqrt{1-\kappa} \hat v_j,% ~~j=1,2,
\end{equation}
where operators $\hat c_j$ are associated with the images of the sources $u_0(x-x_j)$, and $u_0(x)$ is the point spread function~(PSF) of the imaging system, which is assumed to be translationally invariant, and thus $\kappa$ does not depend on the position of the source. Here we assume, without losing generality, that the magnification factor of the imaging system equals 1. Field operators $\hat v_i$ are associated with auxiliary mutually non-orthogonal modes $[ \hat v_i, \hat v_j^\dagger] \ne \delta_{ij}$ in the vacuum state. These auxiliary modes describe losses of light, that unavoidably appear due to the finite aperture of the imaging system and in the general case lead to modes non-orthogonality $[\hat c_j, \hat c_k^\dagger] \ne \delta_{jk}$. %Thus we describe the state of the field in the orthogonal basis of the measurement modes $f_m(x)$, even though this representation is generally infinitely dimensional. 

The measurements are performed in the spatial modes $f_m(x)$ with the corresponding field operators $\hat a_m$, which can be expressed as \cite{karuseichyk_resolving_2022}
\begin{equation} \label{eqn:am1}
     \hat a_m =\sum_j   \left( \int f^*_m (x) u_0(x-x_j) d x \right)~ \hat c_j+\hat v'_m,
\end{equation}
where $\hat v'_m$ are non-normalized field operators of auxiliary vacuum modes. 

In most cases, the PSF is symmetric, i.e. 
$u_0(-x)=u_0(x)$. Thus it is often reasonable to use a symmetric mode basis for the measurement, i.e. perform SPADE in a basis of modes with well-defined parity
\begin{equation} \label{eqn:symmetry}
    f_m(-x)=(-1)^m f_m(x).
\end{equation}
Photon counting in such symmetric bases was shown to be quantum optimal in other scenarios \cite{rehacek_optimal_2017, sorelli_moment-based_2021,karuseichyk_resolving_2022}. In this case Eq.~\eqref{eqn:am1} reduces to
\begin{equation} \label{eqn:am2}
    \hat a_m =  \sum _{j=1}^2  (-1)^{jm} \beta_m ~\hat c_j+\hat v'_m,
\end{equation}
where 
\begin{equation} \label{eqn:beta_HG}
    \beta_m= \int dx~f_m^*(x)~ u_0\left(x-\frac{d}{2}\right).
\end{equation}

% The first order coherency matrix calculated in the measurement basis reads
% \begin{equation}
% \label{eqn:cohA}
%     \mathbf{\Gamma^{(1)\hat a}}%_{mn}=\langle \hat a^\dagger_m \hat a_n \rangle
%     =\vec y_1 \vec y_1^\dagger+\vec y_2 \vec y_2^\dagger.
% \end{equation}
% %Substituting Eq.~\eqref{eqn:cohS} and Eq.~\eqref{eqn:F} we arrive at
%  where 
%  \begin{equation}
%  (\vec y_{1,2})_m=\sqrt{2\kappa N (1 \pm \gamma_0)}  \frac{1\pm (-1)^m e^{i \phi}}{2} \beta_m . %\left( \frac{d}{2} \right)
%  \end{equation}
 
The average photon numbers in the measurement modes reads
 \begin{equation}
     \label{eqn:Ni}
    N_m= \langle \hat N_m \rangle =\langle \hat a^\dagger_m \hat a_m \rangle =\xi_m + \zeta_m 
 \gamma_0 \cos \phi,
 \end{equation}
where 
\begin{equation} \label{eqn:xi}
\xi_m=2\kappa N  \beta^2_m, ~~~
 \zeta_m=2\kappa N  \beta^2_m~(-1)^m.    
\end{equation}
The second term in \eqref{eqn:Ni} accounts for light interference arising from partial coherence. The reader may notice that the average detected photon numbers depend solely on the real part of the degree of mutual coherence, $\operatorname{Re}\gamma = \gamma_0 \cos \phi$. However, as we show in the following, the imaginary part $\operatorname{Im}\gamma$ does influence the photon counting statistics and, consequently, changes the sensitivity of the parameter estimation.  

All the field characteristics after passing through the imaging system contain sources' original brightness $N$ only in combination with the transmissivity factor $\kappa$. Thus important parameter of this problem is the combination $\kappa N$, which defines the brightness of the sources imaged with a given optical system.

The total number of photons detected in $K$ measurement modes equals
\begin{equation} \label{eqn:ND}
    N_{D}=\sum_{m=0}^K N_m=2\kappa N(A+ \delta \gamma_0 \cos \phi).
\end{equation}
Coefficients $A$ and $\delta$ do not depend on the measurement basis if it is full ($K \to \infty$) and symmetric \eqref{eqn:symmetry}:
\begin{equation} \label{eqn:A}
A=\sum_{m=0}^K \beta_m^2 \xrightarrow[K \to \infty]{}1    
\end{equation}
 \begin{equation} \label{eqn:delta}
     \delta=\sum_{m=0}^K \beta_m^2 (-1)^m \xrightarrow[K \to \infty]{} \int u_0\left(x-\frac{d}{2} \right) u_0\left(x+\frac{d}{2} \right) dx.
 \end{equation}
 In this limit ($K \to \infty$), the coefficient $\delta$ depends only on the shape of the PSF $u_0(x)$ and the value of the separation $d$, representing the overlap between images of the sources. In the case of a full measurement basis all the photons, that passed through the imaging system are being detected. Generally, the total number of detected photons $N_D$ \eqref{eqn:ND} depends on the separation $d$ through the overlap $\delta$. One can leverage this dependence to achieve a more accurate estimation of the separation $d$, only provided that the losses are correctly accounted for in the model \cite{tsang_resurgence_2019,kurdzialek_back_2022}. If the theoretical model is defined with an arbitrary loss factor, measuring only one observable does not provide useful information.

\section{Measurement sensitivity}

 To describe the sensitivity of the SPADE measurement, we use the method of moments \cite{kay_fundamentals_1998,gessner_metrological_2019}. This method studies the estimators $\tilde d$ that are based only on the measured sample means $\overline x^{(\mu)}_m$ of the observables~$\hat X_m$. The variance $\Delta^2 \tilde d$ of such estimators is bounded as
\begin{equation} \label{eqn:boundM}
	\Delta^2 \tilde d \ge \frac{1}{\mu M},
\end{equation}
where $\mu$ is the number of measurement repetitions (assumed to be large, so that the central limit theorem can be applied and the statistics of the sample means $\overline x^{(\mu)}_m$ can be considered Gaussian). The sensitivity $M$ is defined as
 \begin{equation}
     \label{eqn:sensitivity}
     M = \sum_{m,n=0}^K \frac{\partial X_m} {\partial d} \left( \mathbf{\Gamma}^{-1} \right)_{mn} \frac{\partial X_n} {\partial d},
 \end{equation}
with $X_m=\langle \hat X_m \rangle$ being the mean values of observables, and $\mathbf{\Gamma}$ the measurement covariance matrix, with elements
 \begin{equation} \label{eqn:Gamma_def}
  \Gamma_{m n}= \langle \hat X_m \hat X_n \rangle -  \langle \hat X_m \rangle \langle \hat X_n \rangle.
 \end{equation} 
The derivation of the bound \eqref{eqn:boundM} and a simple estimator that saturates this bound are presented in Appendix~\ref{appendix_MoM}.
%Moreover, the method of moments presents an optimal estimator \cite{gessner_metrological_2019}. This estimator relies on the linear combination of mean values $N_m$, and asymptotically saturates the bound given by \eqref{eqn:boundM} for a large enough number of measurement repetitions $\mu$.

In the considered scheme, the observables $\hat X_m$ are the photon number operators of the modes $f_m(x)$:
\begin{equation}
    \hat X_m= \hat N_m = \hat a^\dagger_m \hat a_m.
\end{equation}
Exploiting the property of non-displaced Gaussian  states
\begin{multline} 
    \langle \hat a_m^\dagger \hat a_m \hat a_n^\dagger \hat a_n \rangle = 
    \langle \hat a_m^\dagger \hat a_m  \rangle \langle \hat a_n^\dagger \hat a_n \rangle + \\ +
    \langle \hat a_m^\dagger \hat a_n^\dagger \rangle \langle \hat a_m \hat a_n \rangle 
   + \langle \hat a_m^\dagger \hat a_n  \rangle \langle \hat a_m \hat a_n ^\dagger \rangle ,
\end{multline}
and Eq.~\eqref{eqn:am2} we find the photon number covariance matrix%~\eqref{eqn:Gamma_def}
 \begin{multline}
 \label{eqn:Gamma}
    \mathbf \Gamma = \operatorname{diag}\left( \vec Y_1 + \vec Y_2 \right) +  \vec Y_1 \vec Y_1^T + \vec Y_2 \vec Y_2^T + \\ + \frac{1-\gamma_0^2}{2} \sin^2 \phi~~ \vec \zeta \vec \zeta^T,
\end{multline}
where $\operatorname{diag}(\vec x)$ is a diagonal matrix with diagonal elements equals $x_n$ and
  \begin{equation}
 \vec Y_{1,2}=%\left|  (\vec y_{1,2})_n \right|^2&=
 \frac{1 \pm \gamma_0}{2} \left( \vec \xi \pm  \vec \zeta \cos \phi \right),
\end{equation}
with $\vec \xi$ and $\vec \zeta$ defined in Eq.~\eqref{eqn:xi}. 

In the limit of faint sources, $\kappa N \ll 1$, the covariance matrix \eqref{eqn:Gamma} becomes diagonal $\Gamma_{nm}=\delta_{nm} N_m$, and the sensitivity $M$ \eqref{eqn:sensitivity} coincides with the Fisher information. In this case only the real part of the mutual coherence $\operatorname{Re}\gamma = \gamma_0 \cos \phi$ defines the measurement statistics. 
 
In the more general scenario, determining the sensitivity $M$ requires inverting the non-diagonal photon number covariance matrix $\mathbf{\Gamma}$ \eqref{eqn:Gamma}. It is possible to do analytically by applying the Sherman-Morrison formula \cite{sherman_adjustment_1950} three times, but the resulting general expressions are very bulky. Therefore, we decided to focus on several special cases for which the inversion simplifies.
\section{Constant mutual coherence}

 \subsection{Real-valued coherence ($\boldmath{\gamma =\gamma_0}$)}
 First, we consider separation-independent real-valued coherence $\gamma =\gamma_0 \in \mathbb{R}$, i.e. $\phi=0$. For this case, the photon number covariance matrix reduces to:
 \begin{equation}
     \label{eqn:gamma_re}
 \mathbf \Gamma = \operatorname{diag}\left( \vec Y_1 + \vec Y_2 
 %+ N^{dc}(N^{dc}+1) 
 \right) +  \vec Y_1 \vec Y_1^T + \vec Y_2 \vec Y_2^T.
 \end{equation}
%  with 
%  \begin{equation}
%  \vec Y_{1,2}=\frac{1 \pm \gamma_0}{2}\left(\vec \xi \pm \vec \zeta\right).
%  \end{equation}
 The inversion of this matrix gives 
 \begin{equation}  
 \label{eqn:gammaInvRe}
     \Gamma^{-1}_{mn}=
      \delta_{mn} (\xi_n + \gamma_0 \zeta_n)^{-1}-t_{mn},
      %\\     -\frac{(1+(-1)^m)(1+(-1)^n)}{2h_1}      -      \frac{(1-(-1)^{m})(1-(-1)^{n})}     {2h_2}
 \end{equation}
 where
 \begin{equation}
      t_{mn}= \left\{
      \begin{array}{cl}
        2/h_1,  & \text{if}~m \text{ and } n \text{ are both even} \\
        2/h_2,  & \text{if}~m \text{ and } n \text{ are both odd} \\
        0,  & \text{otherwise},
      \end{array}
      \right.
  \end{equation}
  with
 \begin{equation}
 h_{1,2}=2+2\kappa N (1\pm \gamma_0)(A \pm \delta),
 \end{equation}
and $A$ and $\delta$ are defined in Eqs. \eqref{eqn:A} and \eqref{eqn:delta}.

\begin{figure*}[ht]
 \centering
\includegraphics[width=0.99\linewidth]{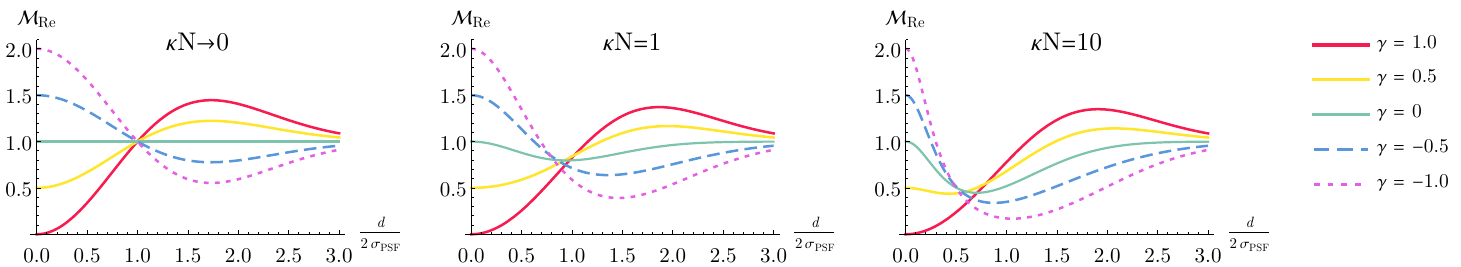}
 \caption{\label{fig:MRe} The normalized sensitivity $\mathcal{M}_{\operatorname{Re}}=  M_{\operatorname{Re}} 4\sigma_{PSF}^2/2\kappa N$ for different real values of the mutual coherence $\gamma$ and numbers of emitted photons $N$ multiplied by the transmissivity $\kappa$ vs. separation $d$.}
 \end{figure*}

  \begin{figure*}[ht] 
 \centering
\includegraphics[width=1.0\linewidth]{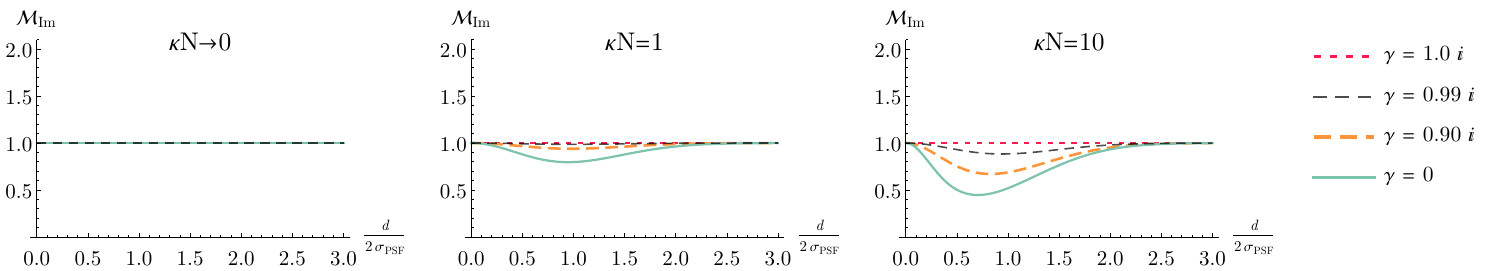}
 \caption{\label{fig:MIm} The normalized sensitivity $\mathcal{M}_{\operatorname{Im}}=M_{\operatorname{Im}} 4\sigma_{PSF}^2/2\kappa N$ for different imaginary values of the mutual coherence $\gamma$ and numbers of emitted photons $N$ multiplied by the transmissivity $\kappa$ vs. separation $d$.}
 \end{figure*}
 
If the mutual coherence $\gamma$ and the brightness $N$ are independent of the separation, i.e. $\partial \gamma_0/\partial d =0$, and  $\partial N/\partial d =0$, the derivative of the measured signal is
\begin{equation}
\label{eqn:Dn}
    \frac{\partial N_m}{\partial d}=2 \frac{N_m}{\beta_m} \frac{\partial \beta_m}{\partial d}.
\end{equation}
Then the normalised sensitivity of the separation estimation is
\begin{multline}
\label{eqn:MRe}  \frac{M_{\operatorname{Re}}}{2\kappa N}= \Delta k^2-\gamma_0 \beta -\\-2 \kappa N (A'+\delta')^2 \left( \frac{(1+\gamma_0)^2}{2h_1}+\frac{(1-\gamma_0)^2}{2h_2} \right), %    1-\frac{\gamma_0}{B} (1-d^2) -2\kappa N B^2 \left(\frac{ d}{2} \right)^2 \left( \frac{(1+\gamma_0)^2}{h_1}+\frac{(1-\gamma_0)^2}{h_2} \right).
\end{multline}
where 
\begin{align}
    \Delta k^2&=4 \sum_{m=0}^K \left(\frac{\partial \beta_m}{\partial d}\right)^2,\\
    \beta &= -4 \sum_{m=0}^K\left(\frac{\partial \beta_m}{\partial d}\right)^2 (-1)^m,\\
    \delta' &= \frac{\partial \delta}{\partial d}=2 \sum_{m=0}^K (-1)^m \beta_m \frac{\partial \beta_m}{\partial d},\\
    A'&=\frac{\partial A}{\partial d}=2 \sum_{m=0}^K \beta_m \frac{\partial \beta_m}{\partial d}.
\end{align}
All these quantities are measurement basis independent for full symmetric bases:
\begin{align}
    \Delta k^2& \xrightarrow[K \to \infty]{} \int \left( \frac{\partial u_0(x)}{\partial x} \right)^2 dx,\\
    \beta & \xrightarrow[K \to \infty]{} \int \frac{\partial u_0(x-d/2)}{\partial x} \frac{\partial u_0(x+d/2)}{\partial x} dx%\\
    %\Omega & \xrightarrow[K \to \infty]{} 0.
\end{align}

In this scenario, the sensitivity expressed in \eqref{eqn:MRe} coincides with the quantum Fisher information \cite{lupo_ultimate_2016, sorelli_quantum_2022}. This implies the possibility of constructing an estimator $\tilde d(\{\overline N^{(\mu)}_m\})$ based solely on the observed sample mean numbers of counts $\overline N^{(\mu)}_m$ in any basis with the parity defined in Eq.~\eqref{eqn:symmetry} and achieving the ultimate resolution limit dictated by the quantum Cramér-Rao bound.

To calculate numerical values of the sensitivity $M$ \eqref{eqn:MRe} we consider the so-called soft-aperture imaging system with Gaussian PSF
\begin{equation} \label{eqn:psf}
    u_0(x)=\left(\frac{1}{2 \pi \sigma_{PSF}^2} \right)^{1/4} \exp \left [{-\frac{x^2}{4 \sigma_{PSF}^2}} \right],
\end{equation}
where $\sigma_{PSF}$ is the PSF width. Then for the full symmetric measurement basis, we have
\begin{align}
    \delta &= \exp \left[ - \frac{d^2}{8\sigma_{PSF}^2}\right],~~
    \Delta k^2 = \frac{1}{4 \sigma_{PSF}^2} \\
    \beta &= \frac{\sigma_{PSF}^2-(d/2)^2}{4\sigma_{PSF}^4} \exp \left[ - \frac{d^2}{8\sigma_{PSF}^2}\right] .
\end{align}
 The sensitivity calculated with Eq.~\eqref{eqn:MRe} is shown in Fig.~\ref{fig:MRe}. The first plot corresponds to a small number of emitted photons and matches other results obtained in the same limit \cite{tsang_resurgence_2019,kurdzialek_back_2022}. With an increasing number of photons, there is a reduction in the sensitivity for the separations $d \approx \sigma_{PSF}$. This reduction is also present in the QFI and occurs due to the quadratic term of the noise in the thermal statistics \cite{lupo_ultimate_2016}. However, as we show in the next section, this drop does not occur even for bright thermal states, if their mutual coherence degree $\gamma$ is not a real number. 

 \subsection{Imaginary coherence ( $\boldmath{\gamma =i \gamma_0}$) }
 Another interesting example that we consider is that of a purely imaginary degree of coherence $\gamma =i \gamma_0 \in \mathbb{I}$, which corresponds to a relative phase $\phi=\pi/2$ between the sources. In this case, the measured mean photon numbers $N_m$ do not contain an interference term, i.e. it is the same as for a pair of incoherent sources. However, the presence of coherence does influence the covariance matrix
  \begin{equation}
 \label{eqn:GammaIm}
     \mathbf \Gamma = 
     \operatorname{diag}\left( \vec \xi \right)  + \frac{1+\gamma_0^2}{2}~ \vec \xi \vec \xi^T   +  \frac{1-\gamma_0^2}{2}~ \vec \zeta \vec \zeta^T,
 \end{equation}
which only depends on $\gamma_0^2$ and thus is not affected by its sign. The inverse of this matrix reads
\begin{equation}
     \Gamma^{-1}_{mn}=
     %\frac{1}{2} 
     \delta_{mn} \xi_n- 
     \frac{(-1)^{m+n} \eta_1 + \eta_2 -2\kappa N \delta ((-1)^n+(-1)^m)}
     {\eta_1 \eta_2 - (2\kappa N \delta)^2},
 \end{equation}
where $\eta_{1,2}=2 \kappa N A + 2(1\pm \gamma_0^2)^{-1}$. In this case, the normalized sensitivity takes the form 
 \begin{multline}
\label{eqn:Mim}
    \frac{M_{\operatorname{Im}}}{2\kappa N}= \Delta k^2 
    -2 \kappa N 
    \frac{ \eta_1 (\delta')^2 + \eta_2 (A')^2 -2\kappa N \delta A' \delta'  }{\eta_1 \eta_2 - (2\kappa N \delta)^2 }.
\end{multline}
It is plotted in Fig. \ref{fig:MIm} for different values of the mutual coherence and sources' intensities.

 One can see that in the limit of faint sources the sensitivity
\begin{equation}
      \frac{M_{\operatorname{Im}}}{2\kappa N} \xrightarrow{\kappa N \to 0} \Delta k^2
\end{equation}
is the same for any imaginary degree of mutual coherence, including the case of incoherent sources.
 
 However, for bright incoherent thermal sources the sensitivity per photon drops significantly in the sub-Rayleigh region $d<\sigma_{PSF}$, while for sources with imaginary mutual coherence, this drop is smaller or doesn't occur at all for perfect coherence $\gamma=i$. Generally, if imaginary coherence is close to the complex unity % i.e. $1-\gamma_0 \ll 1$,
 then the sensitivity can be expressed as
\begin{equation}
    \frac{M_{\operatorname{Im}}}{2\kappa N} =\Delta k^2 - 2 \kappa N (\delta')^2 ~ (1-\gamma_0) +O((1-\gamma_0)^2).
\end{equation}
This means that the presence of correlations between sources can improve the sensitivity of the separation estimation even in cases when no interference is visible in the mean values $N_m$. % (due to the relative phase $\phi=\pi/2$). 
One can use a simple intuition to interpret this observation: two incoherent thermal sources do not share any correlations, while mutually coherent sources are correlated in the number of photons, even if they do not interfere due to the mutual phase $\phi = \pi/2$. Utilizing these correlations allows to cancel part of the intensity noise in the measurement modes, resulting in a more precise estimation of the separation.

\section{Parameter-dependent coherence} 
Now let us consider the situation when the mutual coherence $\gamma$ depends on the separation  $d$ of the sources. Below we will discuss in details physical examples of such systems, but first, we analyze the general case of real-valued separation dependent mutual coherence $\gamma(d)=\gamma_0(d) \in \mathbb{R}$.

This dependence leads to an additional term in the derivative 
\begin{equation}
\label{eqn:Dngamma}
    \frac{\partial N_m}{\partial d}=2 \frac{N_m}{\beta_m} \frac{\partial \beta_m}{\partial d}+ \frac{\partial \gamma_0}{\partial d} \zeta_m. %\cos \phi
\end{equation}
%The inverse of the photon number covariance matrix for this case is given by \eqref{eqn:gammaInvRe} and 
and in the sensitivity 
\begin{equation} \label{eqn:Mgamma}
    M_{\gamma_0(d)}=M_{\operatorname{Re}} + \Delta M_{\gamma_0(d)},
\end{equation}
 where
\begin{multline}
\label{eqn:dMgamma}
    \frac{\Delta M_{\gamma_0(d)}}{2\kappa N}=2\gamma_0'
    \Bigg[\delta' \left( \frac{1}{h_1}+\frac{1}{h_2} \right)+A' \left(\frac{1}{h_1} -\frac{1}{h_2} \right)\Bigg] +\\
    +\left(\gamma_0'\right)^2
    \Bigg[\frac{A-\delta \gamma_0}{1-\gamma_0^2} - 2\kappa N \left(\frac{(A+\delta)^2}{2h_1}+\frac{(A-\delta)^2}{2h_2} \right) \Bigg],
\end{multline}
and $M_{\operatorname{Re}}$ is defined in Eq.~\eqref{eqn:MRe}.

The additional sensitivity $\Delta M_{\gamma_0(d)}$ originates from the fact that in the case of separation-dependent coherence, the measurement results change faster with changing separation, and one can do estimation with higher precision. The extra sensitivity $\Delta M_{\gamma_0(d)}$ typically has maxima around the maxima of $\gamma_0'$, i.e. in the regions of fast-changing mutual coherence.

In the limit of faint sources, the additional sensitivity takes the form
\begin{equation} 
    \frac{\Delta M_{\gamma_0(d)}}{2\kappa N} \xrightarrow{\kappa N \to 0} ~2 \gamma_0' \delta' + (\gamma_0')^2 \frac{A-\gamma_0 \delta}{1-\gamma_0^2}.
\end{equation}
This expression is valid for an arbitrary complex value of the mutual coherence $\gamma$ if one replaces $\gamma_0$ with $\operatorname{Re} \gamma$. 

\subsection{Finite coherence width of the illumination}
\begin{figure}[h]
 \centering
\includegraphics[width=.75 \linewidth]{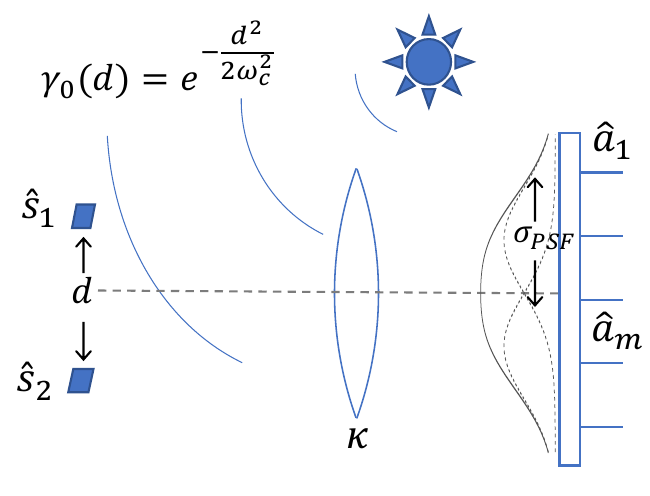}
 \caption{\label{fig:common} Objects reflect the light from the common thermal illumination source with finite coherence width $\omega_c$.}
 \end{figure}

In the first example, we investigate the estimation of separation between two reflective objects. When illuminated with light of finite coherence width, the mutual coherence of the reflected light becomes dependent on the separation between the reflectors (see Fig. \ref{fig:common}). 

In many cases, the transverse coherence of the illuminating light can be effectively approximated by a Gaussian function. In such scenarios, the field reflected by two small reflectors exhibits mutual coherence
\begin{equation}
\label{eqn:c(d)}
    \gamma_0(d)= %c_1
    \operatorname{exp}\left[-\frac{d^2}{2 \omega_c^2} \right],
\end{equation}
where the coherence width $\omega_c$ can be adjusted by changing the optical parameters of the illumination system. % The coherence time, and consequently the emitted number of photons per coherence time, can be modified by adjusting the speed of the disk rotation. 
%Note that in practical situations the number of coherently reflected photons often can be large.

\begin{figure}[t]
 \centering
 \includegraphics[width=.69\linewidth]{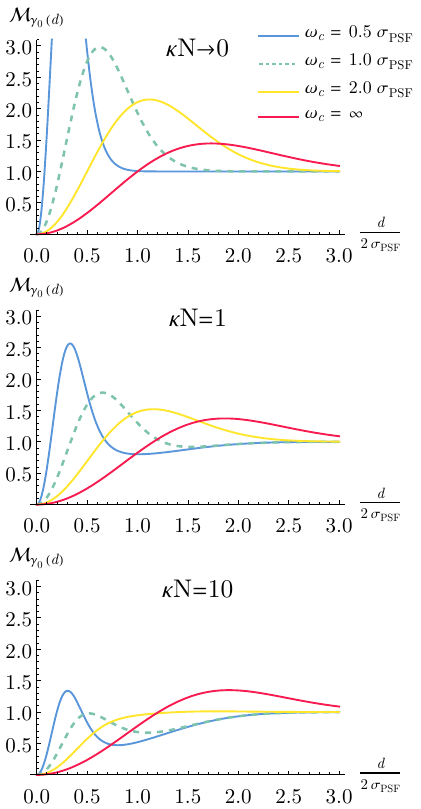}
 \caption{\label{fig:Mgamma} Normalized sensitivity $\mathcal{M}_{\gamma_0(d)}=M_{\gamma_0(d)} 4\sigma_{PSF}^2 /2\kappa N$  in case of the separation-dependent mutual coherence with Gaussian profile \eqref{eqn:c(d)}.}
 \end{figure}

Using the Gaussian PSF \eqref{eqn:psf}, we calculate the full sensitivity  \eqref{eqn:Mgamma} of the reflectors' separation estimation (Fig.~\ref{fig:Mgamma}). The red lines on the plots correspond to the spatially coherent illumination. One can see that separation-dependent coherence, compared to fully coherent case, results in significantly higher sensitivity for the separations close to the coherence width of the illumination source $d \approx \omega_c$, thus it is desirable to use illumination with coherence width of the order of the measured separations. A similar increase in the resolution, when the coherence width of the source matches the size of the features of the studied object, was observed earlier experimentally and numerically for the case of quantum imaging with pseudo-thermal light \cite{mikhalychev_efficiently_2019}.  

The coherence width of the light expands as it passes through a diffractive imaging system. Thus, if the same optical system (with PSF width $\sigma_{PSF}$) is used for the imaging and the illumination of the object, the minimal achievable coherence width of the illumination is $\omega_c \ge 2 \sigma_{PSF}$. %To achieve a shorter coherence width one needs to use better optical system.
This case was studied in Ref. \cite{wang_quantum-limited_2023} in the faint source limit, where an analogous increase of the SPADE sensitivity was demonstrated for the separations around the Rayleigh limit (in comparison to spatially coherent illumination). However, as we show in Fig. \ref{fig:Mgamma}, for brighter sources this increase is much less noticeable. In the extreme case of very bright sources, additional sensitivity per photon \eqref{eqn:dMgamma} vanishes
 \begin{equation}
 \frac{\Delta M_{\gamma_0(d)}}{2\kappa N} \xrightarrow[N \to \infty]{} 0 .
 \end{equation}
  This is important to take into account for practical imaging with pseudo-thermal sources, since this scenario often does not fit into the faint sources approximation. Therefore, it is crucial to aim for the utilization of an illumination source with a short coherence time to ensure that the number of photons $N$ reflected by each imaged object per coherence time is small.
 
 One could achieve better sensitivity in the sub-Rayleigh regime via a smaller coherence width, which is achievable in two possible scenarios: using an independent illumination scheme with narrower PSF (for example, if the illumination source is located closer to objects than the detection apparatus) or in the case of sources' mutual coherence originating from their interaction, which is considered in the next section.
 
\subsection{Interacting emitters}
\begin{figure}[h]
 \centering
\includegraphics[width=.76\linewidth]{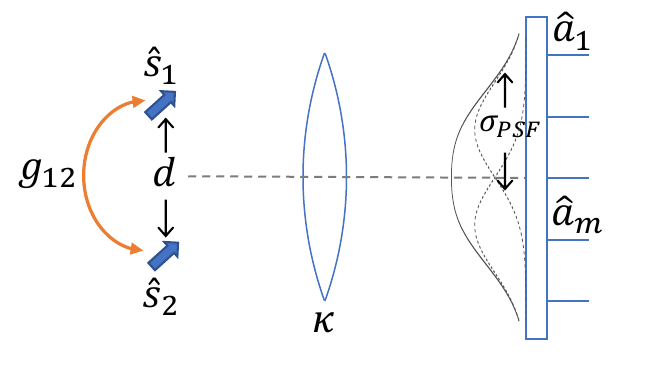}
 \caption{\label{fig:interaction} Dipoles emit partially coherent light due to the interaction.}
\end{figure}

Another possible scenario where the mutual coherence of the emitted light depends on the distance between the sources is the case of interacting emitters (see Fig. \ref{fig:interaction}). As an example, we consider two identical, dipole-dipole interacting two-level systems prepared initially in their excited states. In this scenario, each dipole emits precisely 1 photon during the decay of the excited state. For simplicity, we assume that the dipole moments are parallel to each other and the line connecting them is orthogonal to the main optical axes of the imaging system.

The time evolution of the quantum state of the two coupled dipoles in the interaction picture can be described by the following master equation \cite{lehmberg_radiation_1970,agarwal_quantum_1974}
\begin{multline} \label{eqn:dipole_dynamic}
    \frac{d}{dt} \hat \rho = -i f_{12} [\hat \sigma_1^+ \hat \sigma_2^- +h.c.,\hat \rho] ~+\\
    \frac{1}{2} \sum_{j,l=1}^2 g_{jl}\left( 2 \hat \sigma_j^- \hat \rho \hat \sigma_l^+ - \hat \sigma_j^+ \hat \sigma_l^- \hat \rho  -
     \hat \rho  \hat \sigma_j^+ \hat \sigma_l^-
    \right)+\\
  {\sum_{j=1}^2 \eta_j \left( 2 \hat \sigma_j^+ \hat \sigma_j^- \hat \rho \hat \sigma_j^+ \hat \sigma_j^- - \hat \sigma_j^+ \hat \sigma_j^- \hat \rho  -
     \hat \rho  \hat \sigma_j^+ \hat \sigma_j^-
    \right)},
\end{multline}
where $\hat \sigma^\pm_j$ are the dipoles' transition operators. The first term in Eq.~\eqref{eqn:dipole_dynamic} is proportional to the collective frequency shift $f_{12}$ of the energy levels \cite{agarwal_quantum_1974,james_frequency_1993}. %Note that this unitary term does not affect the sensitivity of our frequency-blind measurements of the fluorescence intensity and thus the constant $f_{12}$ does not appear in further calculations. 
The second term describes the spontaneous emission of the individual emitters ($j=l$) and their collective radiative decay ($j\neq l$). In the case of identical dipoles,
\begin{gather} \label{eqn:g12}
     g_{11}=g_{22}= 1/\tau, \\ 
     g_{12}=\frac{3}{2} \frac{1}{\tau} \left ( \frac{\sin z}{z}+\frac{\cos z}{z^2}-\frac{\sin z}{z^3} \right),
\end{gather}
where $\tau$ is the natural lifetime of the excited state of an individual dipole, and $z=2 \pi d/\lambda$, with $\lambda$ the wavelength of the dipole transition.  {The final term in the master equation accounts for the dephasing of the dipoles with the rates $\eta_i$ ($i=1,2$), due to their different local environments.}

The radiation of the dipoles is described in the transient temporal modes $\hat s_{1,2}(t)$, of duration $\Delta t \ll \tau$, centered around time $t$. For these modes, the  time-dependent coherency matrix can be expressed through the atomic dipole correlators as
\begin{equation}
    \label{eqn:coh2}
    \Gamma^{(1)\hat s(t)}_{j k} =  \langle \hat \sigma^+_j \hat \sigma^-_k \rangle \frac{\Delta t}{\tau}.
\end{equation}
%Its analytical form is presented in Appendix \ref{appendix1}. 
The coherency matrix of the excited modes remains unaffected by the collective shifts $f_{12}$. However, as discussed in Appendix \ref{appendix1}, the emission spectrum is influenced by $f_{12}$. Consequently, the measurement results and the sensitivity of the measurement are independent of $f_{12}$ exclusively in the considered case of frequency-blind measurements. 

Cross-correlation of the atomic operators $\langle \hat \sigma^+_1 \hat \sigma^-_2 \rangle$ vanishes if the collective decay rate $g_{12}=0$. The crucial role of these correlations in the emergence of quantum coherence was previously discussed in other contexts \cite{shatokhin_coherence_2018,ames_theory_2022}. 

Both the unitless emission rate $\dot{N}=N \tau/\Delta t=\Gamma^{(1)\hat s(t)}_{1 1} \tau/\Delta t$ and the degree of mutual coherence $\gamma=\Gamma^{(1)\hat s}_{1 2}/\Gamma^{(1)\hat s}_{1 1}$ depend on both time and separation, however, we will not explicitly indicate it in our notations. 

\begin{figure}[h]
    \centering
    \vspace{8pt}
    \includegraphics[width=.67 \linewidth]{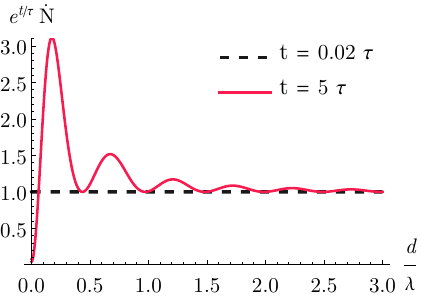}\\ 
    \includegraphics[width=.71 \linewidth]{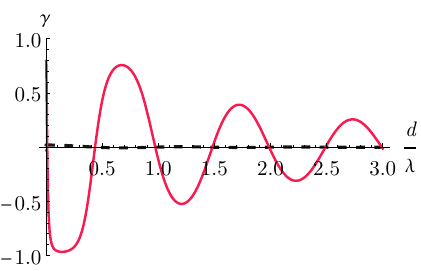}
    \caption{Properties of the dipoles radiation:  the normalised emission rate $\dot{N}$ and the degree of mutual coherence $\gamma$ vs. separation $d$. The dashed line corresponds to the time $t=0.02 \tau$, solid line $t=5 \tau$. No dephasing ($\eta=0$). }
    \label{fig:dipole_radiation}
\end{figure}

{\subsubsection{Model without dephasing ($\eta_1=\eta_2=0$)}
First, we examine the evolution of the dipoles' state without taking into account the dephasing process. The analytical solution for the dipole correlators in this scenario is presented in Appendix \ref{appendix1}. The associated emission rate $\dot{N}$ and the degree of mutual coherence $\gamma$ for different time instances $t$ are depicted in Fig.~\ref{fig:dipole_radiation} as functions of the separation $d$ normalized by the wavelength~$\lambda$.}
%We display plots of these variables for various time moments $t$ as functions of the separation $d$ normalized with wavelength $\lambda$ in Fig.~\ref{fig:dipole_radiation}.

One can see that at an early stage of the emission ($t=0.02 \tau$) the dipoles fluoresce almost independently, with the emission rate close to that of an individual dipole, $\dot{N} \approx e^{-t/\tau}$, and with almost no coherence, $\gamma \approx 0$. However, after some time, the dipole-dipole interaction creates correlations and the difference with the individual dipole emission becomes evident in both the emission rate \cite{devoe_observation_1996} and the mutual coherence. 
Crucially, this difference depends on the separation between the dipoles, which allows for a measurement of the separation with higher accuracy. Without accounting for the decoherence process, the mutual coherence exhibits a particularly strong dependence on separation during the late emission stages ($t>\tau$) for separations $d\approx n \frac{\lambda}{2}, n \in \mathbb{N}$. Therefore, we can expect heightened sensitivity in estimating separation around these points.
Note that due to the geometrical symmetry of the problem the mutual coherence $\gamma$ stays real at any moment, i.e. $\gamma (t)  \in \mathbb{R}~\forall t$. However, all the conclusions below can be generalized to the case of the complex degree of mutual coherence by replacing $\gamma_0$ with $\operatorname{Re}~\gamma$.

The mean number of photons emitted by each dipole during a short time interval $\Delta t\ll\tau$ is low ($N \ll 1$). Furthermore, the imaging system is characterized by high losses, $\kappa \ll 1$. Therefore, irrespectively of the statistics of the sources, the detection statistics can be described by a Poisson distribution which also approximates that of faint thermal sources discussed earlier. Besides, the radiation emitted by the two dipoles in the far field represents a superposition of two spherical waves and therefore is locally equivalent to the radiation emitted by two-point sources (see Appendix~\ref{appendix1}). Thus, in the paraxial approximation, the model developed in the previous sections can be applied to the problem of resolving dipoles. 

The only modification required to Eq.~\eqref{eqn:Dngamma} is the addition of an extra term, associated with the separation-dependent emission rate:
\begin{equation}
\label{eqn:DngammaN}
    \frac{\partial N_m}{\partial d}=2 \frac{N_m}{\beta_m} \frac{\partial \beta_m}{\partial d}+ \frac{\partial \gamma_0}{\partial d} \zeta_m \cos \phi+\frac{N_m}{N}\frac{\partial N}{\partial d}.
\end{equation}

\begin{figure}%[b]
 \centering
     \begin{tabular}{c}
\includegraphics[width=.67\linewidth]{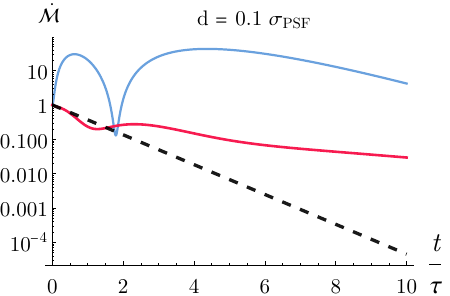} 
     \end{tabular}
 \begin{tabular}{c}
   \includegraphics[width=.28\linewidth]{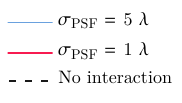} \\
\end{tabular}
 \caption{\label{fig:Mdipole_t} Normalized sensitivity rate $\dot{\mathcal{M}}$~\eqref{eqn:M_dot} of two dipoles' separation estimation vs. time $t$. The light blue line corresponds to $\sigma_{PSF}=5\lambda$, the red line to $\sigma_{PSF}=\lambda$, and the dashed line to non-interacting dipoles.}
 \end{figure}

\begin{figure*}%[t]
 \centering
\includegraphics[width=1\linewidth]{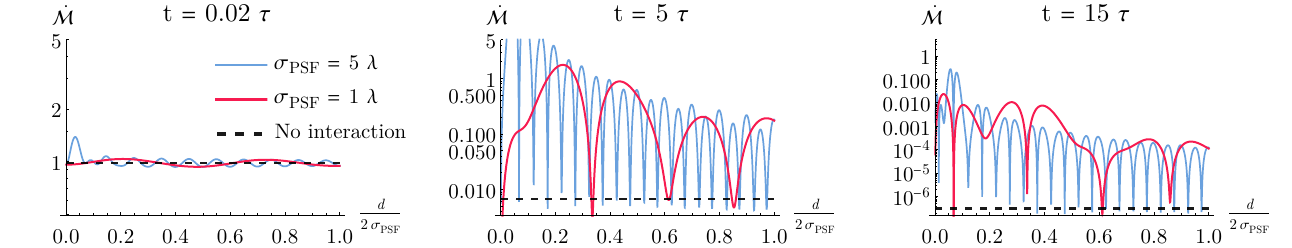}
 \caption{\label{fig:Mdipole} Normalized sensitivity rate $\dot{\mathcal{M}}$~\eqref{eqn:M_dot} of two dipoles' separation estimation vs. separation $d$. The light blue line corresponds to $\sigma_{PSF}=5\lambda$, the red line to $\sigma_{PSF}=\lambda$, and the dashed line to non-interacting dipoles. No dephasing ($\eta=0$).} 
 \end{figure*}

 This extra-dependence results in an additional term $ \Delta M_{N(d)}$ in the sensitivity. In the limit $\kappa N=\kappa \dot{N} \Delta t/\tau \ll 1$ it takes the form
\begin{equation}
    \Delta M_{N(d)} = \frac{2 \kappa N'}{N} \Big[ N'
    (A+\gamma \delta)+2N(A+\gamma \delta)'\Big],
\end{equation}
where $X'=\partial X/\partial d$. Combining all the contributions we get the normalized sensitivity per unit time in the form
\begin{equation} \label{eqn:M_dot}
     \dot{\mathcal{M}} = \frac{4\sigma_{PSF}^2 }{2 \kappa} \frac{\tau}{\Delta t} \left( M_{\operatorname{Re}} + \Delta M_{\gamma_0(d)} + \Delta M_{N(d)}\right).
      \end{equation}
This quantity is plotted in Fig. \ref{fig:Mdipole_t} and \ref{fig:Mdipole}. 

As one can expect, at an early stage of emission the problem resembles the resolving of two incoherent sources (i.e. non-interacting emitters). In the context of independent emitters, the sensitivity per unit time is directly proportional to the emission rate, exhibiting an exponential decrease, as indicated by $\dot{N} \approx e^{-t/\tau}$ (depicted by the dashed black lines in Fig. \ref{fig:Mdipole_t} and \ref{fig:Mdipole}). On the contrary, in the case of interacting emitters, the sensitivity rate may even increase over time due to the cumulative effects of interaction. The specific values of the sensitivity depend on the ratio between the separation $d$, PSF width $\sigma_{PSF}$ and the wavelength $\lambda$. We present plots for the cases of $\sigma_{PSF}=\lambda$ and $\sigma_{PSF}=5\lambda$. In the paraxial approximation, the PSF width of the imaging system typically significantly exceeds the wavelength, i.e. $\sigma_{PSF} > \lambda$. Therefore, the case $\sigma_{PSF}=\lambda$ stretches the boundaries of our model. Nevertheless, we investigate this case to explore the model's limits and offer an illustrative example that is easy to analyze.

In Fig. \ref{fig:Mdipole}, one can observe a pronounced enhancement in sensitivity within regions where the emission characteristics (the mutual coherence and the emission rate) exhibit rapid changes with a change in separation. Meanwhile, the sensitivity remains low around the extrema of the emission characteristics. In typical cases, the sensitivity for interacting dipoles is significantly larger and decaying significantly slower compared to non-interacting dipoles. Even after more than 15 lifetimes of the excited state, when the probability of photon detection is very low, the average sensitivity rate remains notably high due to the substantial amount of information in ``late" detection events. 

It is important to note that the moment-based sensitivity provides a limit for a local estimation strategy. The presence of multiple narrow peaks in the sensitivity plot signals the potential degeneracy of the estimator. Therefore, when dealing with a low-resolution optical system ($\sigma_{PSF} \gg \lambda$), one may require an increased number of measurement repetitions to reach the saturation of the bound \eqref{eqn:boundM}.

\begin{figure}[h]
\centering
\includegraphics[width=.75\linewidth]{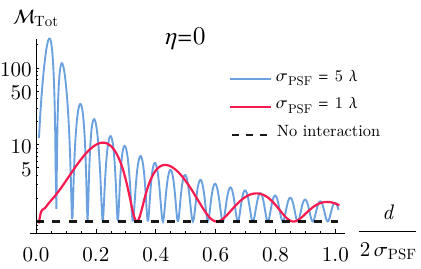} 
 \caption{\label{fig:Mdipole_d} Normalized total sensitivity $\mathcal{M}_{Tot}$~ \eqref{eqn:M_tot} of two dipoles' separation estimation vs. separation $d$. The light blue line --- $\sigma_{PSF}=5\lambda$, the red line --- $\sigma_{PSF}=\lambda$, the dashed line --- non-interacting dipoles. 
 No dephasing ($\eta=0$).}
 \end{figure}

In general, the interaction between the emitters induces an entanglement between them \cite{tanas_entangling_2004} and creates temporal correlations of the emitted light \cite{peshko_quantum_2019}. However, if the losses in the imaging system are large $\kappa \ll 1$, the detection events for different time intervals can be considered independent. The sensitivity of independent detection events is additive thus the total sensitivity can be calculated as
\begin{equation} \label{eqn:M_tot}
    \mathcal{M}_{Tot}= \frac{1}{\tau} \int_0 ^\infty \dot{\mathcal{M}}~dt,
\end{equation}
where $\dot{\mathcal{M}}$ is defined in \eqref{eqn:M_dot}.
The result of this calculation is shown in Fig. \ref{fig:Mdipole_d}. One can see that the interaction effects can increase the total sensitivity of the separation estimation by several orders of magnitude around the points $d\approx n \frac{\lambda}{2}, n \in \mathbb{N}$, where the mutual coherence of the emission strongly depends on the separation, if the dephasing effect is not present. In between these points, for the separations around $d\approx \left(n - \frac{1}{2} \right) \frac{\lambda}{2}, n \in \mathbb{N}$, both the emission rate and mutual coherence exhibit extrema. Consequently, the sensitivity is not heightened by the separation-dependent emission characteristics around these specific points. 

{\subsubsection{Model with dephasing ($\eta_1,\,~\eta_2\neq 0$)}
In realistic scenarios, the process of building up coherence through the dipole-dipole interactions competes with the coherence loss caused by dipoles' individual dephasing. 
The master equation \eqref{eqn:dipole_dynamic} with dephasing can be solved analytically, however the solution is too long and cumbersome to be presented within this paper. 
Nevertheless, it demonstrates that the coherency matrix for our system only depends on the average dephasing rate $\eta=(\eta_1+\eta_2)/2$ of the dipoles. In the following, we discuss the impact of dephasing for different values of $\eta$. Figure~\ref{fig:dipole_radiation_withDecoherence} illustrates the emission characteristics (the emission rate and the mutual coherence) at the time instance $t=5 \tau$. In the weak dephasing regime ($\eta=0.2/\tau$), where the dephasing time is substantially longer than the natural lifetime of the excited state $\tau$, the effect of dephasing has a limited influence on the emission process. On the other hand, when the dephasing process is faster than the emission ($\eta = 2.0/\tau$), one can observe a significant decrease in the mutual coherence $\gamma$. 

\begin{figure}[b]
    \centering
    \vspace{8pt}
    \includegraphics[width=.63 \linewidth]{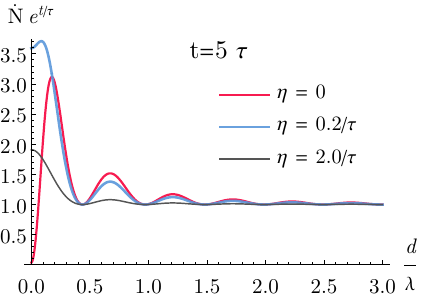}\\ 
    \includegraphics[width=.65 \linewidth]{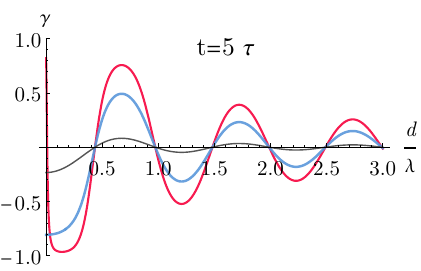}
    \caption{{Emission characteristics:  normalized emission rate $\dot{N}$ and degree of mutual coherence $\gamma$ at the time moment $t=5 \tau$. The red line corresponds to the dephasing-free model $\eta=0$, the light blue line -- weak dephasing $\eta=0.2/\tau$, the grey line -- strong dephasing $\eta=2.0/\tau$. }}
    \label{fig:dipole_radiation_withDecoherence}
\end{figure}

As expected, the reduction in emission coherence due to the dephasing effect leads to a weaker sensitivity boost. Fig. \ref{fig:Mdipole_d_withDecoherence} illustrates the total sensitivity of the separation estimation defined in \eqref{eqn:M_tot}. The top plot highlights the robustness of the considered scheme to weak dephasing, where the sensitivity is only minimally affected, compared to the dephasing-free case, shown in Fig. \ref{fig:Mdipole_d}. Conversely, the bottom plot in Fig. \ref{fig:Mdipole_d_withDecoherence} reveals that strong dephasing significantly diminishes the sensitivity boost arising from dipole interaction. However, even in this scenario, the sensitivity can be several times higher compared to non-interacting dipole emission.
\begin{figure}[h]
\centering
\includegraphics[width=.7\linewidth]{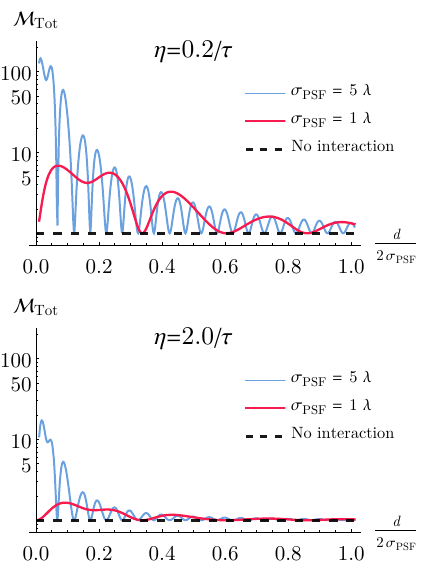} 
 \caption{{Normalized total sensitivity $\mathcal{M}_{Tot}$~ \eqref{eqn:M_tot} for the model with dephasing. Top inset -- weak dephasing $\eta=0.2/\tau$, bottom inset -- strong dephasing $\eta=2.0/\tau$. The light blue line --- $\sigma_{PSF}=5\lambda$, the red line --- $\sigma_{PSF}=\lambda$, the dashed line --- non-interacting dipoles.}}
\label{fig:Mdipole_d_withDecoherence}
 \end{figure}}

{\subsubsection{Model with detection noise}
Another factor that can potentially significantly reduce sensitivity is the presence of detection noise. If the detection system produces dark counts or detects background light, then late rare detection events may be dominated by these false counts. To address this effect, we modify the detection covariance matrix \eqref{eqn:Gamma} as 
\begin{equation} \label{eqn:Gamma_DC}
    \mathbf{\Gamma}_{\text{DC}} = \mathbf{\Gamma} +  N_{\text{DC}} \mathbf{I},
\end{equation}
where $\mathbf{I}$ stands for the identity matrix, and $N_{\text{DC}}$ is the average number of dark counts per detection mode. Here, the dark counts are assumed to have Poisson statistics, being uncorrelated from each other and the measured signal.

Using the covariance matrix of noisy detection \eqref{eqn:Gamma_DC}, we numerically calculate the sensitivity \eqref{eqn:sensitivity}, taking into account the separation-dependent emission rate \eqref{eqn:N_dot_appendix} and mutual coherence \eqref{eq:Ng} of dipoles. We consider detection in the first four Hermite-Gauss modes and express the rate of dark counts relative to the rate of real counts at the early stages $\dot{N}_0 = 2\kappa$. Note, that the rate of real detection events exponentially decreases with time, while the rate of dark counts remains constant.

The sensitivity rate for noisy detection is presented in Fig. \ref{fig:Mdipole_t_DC}. Here we consider strong detection noise, with the dark count rate being 50\% of the initial rate of real counts  $\dot{N}_{\text{DC}}=0.5\dot{N}_0$. Comparing this plot to the ideal case (Fig. \ref{fig:Mdipole_t}), one can observe a significant decrease in the sensitivity rate for late detection events. After $t=5 \tau$ one obtains almost no information about the separation of dipoles. This decrease occurs because the dark count rate is much higher than the rate of real counts for these late stages. As a result, in this case, most of the useful information comes from the early and intermediate radiation stages ($t<5 \tau$).

\begin{figure}[h]
 \centering
     \begin{tabular}{c}
\includegraphics[width=.67\linewidth]{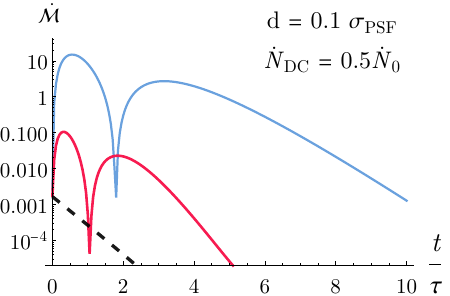} 
     \end{tabular}
 \begin{tabular}{c}
   \includegraphics[width=.28\linewidth]{Mdipole_leg.pdf} \\
\end{tabular}
 \caption{\label{fig:Mdipole_t_DC} Normalized sensitivity rate $\dot{\mathcal{M}}$ of two dipoles' separation estimation vs. time $t$ in case of noisy detection $\dot N_{\text{DC}}=0.5\dot N_0$. %The light blue line --- $\sigma_{PSF}=5\lambda$, the red line --- $\sigma_{PSF}=\lambda$, the dashed line --- non-interacting dipoles.
 }
 \end{figure}
 
\begin{figure}[h]
\centering
\includegraphics[width=.73\linewidth]{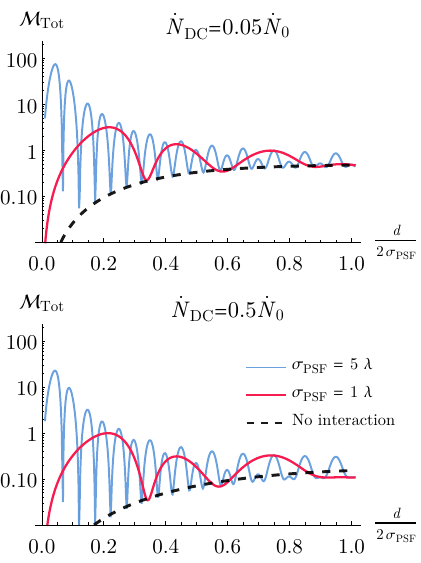} \\
 \caption{{Normalized total sensitivity $\mathcal{M}_{Tot}$ vs. separation $d$ for the model with detection noise. Top inset --- low rate of the dark counts, bottom inset --- intense noise.}}
\label{fig:Mdipole_d_DC}
 \end{figure}
}

Integrating the sensitivity rate over time we obtain the total sensitivity \eqref{eqn:M_tot} for noisy detection and present it in Fig. \ref{fig:Mdipole_d_DC}. As expected, compared to the ideal case (Fig. \ref{fig:Mdipole_d}), the sensitivity is suppressed in the presence of the detection noise. However, this is also the case for non-interacting dipoles. Thus, one can still see a significant increase in the sensitivity (two to three orders of magnitude for small separations) coming from the dipole interaction, even for relatively intense detection noise $\dot{N}_{\text{DC}}=0.5\dot{N}_0$. 

\vspace{13pt}
Curiously, in all the considered cases the sensitivity boost is more pronounced for smaller separations, where the interaction between the emitters is stronger. As a result, in the context of single-parameter estimation, one can achieve higher precision in estimating smaller separations compared to larger ones.

Finally, we note that in the case of frequency-resolved measurements, the collective frequency shift $f_{12}$ may come into play, such that one may be able to locally estimate the dipoles' separation via the emission spectrum. Therefore, by integrating spectral and spatial approaches, the sensitivity could be further enhanced by introducing frequency-resolving detection into our scheme.

\section{Conclusion}
We have analyzed the problem of resolving partially coherent thermal sources with SPADE measurement. To estimate the sensitivity of a separation estimation, we utilized the method of moments, which allowed us to make no assumptions about the brightness of the sources. We have found analytical expressions for the sensitivity including the cases of separation-dependent mutual coherence and emission rate. We studied two specific examples of separation-dependent coherence: a reflection of light coming from a finite-coherence-width illumination source and creating mutual coherence due to the interaction of the emitters. In both cases, we demonstrate the possibility of a significant boost in the separation estimation sensitivity due to the additional mechanism of the parameter encoding to the problem. Our analysis shows that for efficient resolving of the reflecting objects one needs to use an illumination source with a narrow coherence width (on the order of the reflectors' separation) and a short coherence time. This ensures the full advantage of the separation-dependent coherence of the reflected light. Examining the interacting emitters, we demonstrate that the sensitivity of separation estimation can be increased by several orders of magnitude compared to independent emitters. This enhancement arises from the separation-dependent mutual coherence and emission rates of the interacting dipoles. The effect remains robust in the presence of weak dephasing of dipoles. Although strong dephasing decreases the observed resolution boost, it does not eliminate it entirely. We have also demonstrated that the presence of detection noise reduces sensitivity but preserves the relative boost caused by the interaction of dipoles.
\vspace{10pt}
\section*{ACKNOWLEDGMENTS}
I.K. acknowledges support from the PAUSE program. This project has received funding from the European Defense Fund (EDF) under grant agreement EDF-2021-DIS-RDIS-ADEQUADE (n°101103417). Funded by the European Union. Views and opinions expressed are however those of the author(s) only and do not necessarily reflect those of the European Union. Neither the European Union nor the granting authority can be held responsible for them. This work was carried out during the tenure of an ERCIM ‘Alain Bensoussan’ Fellowship Programme.

\appendix
\section{The method of moments} \label{appendix_MoM}
Let us study the estimation of a parameter $\theta$ from measuring a set of observables $\vec {\hat X}=(\hat X_1, \hat X_2,...)^T$. 
More specifically, we limit ourselves to estimators $\tilde \theta$ that are based on the sample means $\overline{x}_m^{(\mu)}$ of these observables, obtained from $\mu$ repetitions of the measurements:
\begin{equation} \label{eqn:generalMoMsetimator}
    \tilde \theta=f(\overline{x}^{(\mu)}_1,\overline{x}^{(\mu)}_2,...).
\end{equation}

Following the central limit theorem, for sufficiently large statistics $\mu \gg 1$ the sample means $\overline{x}^{(\mu)}_m$ follow the multivariate Gaussian distribution $\mathcal{N}(\langle \vec{\hat X} \rangle, \frac{1}{\mu} \mathbf{\Gamma})$.  The Fisher information in the case of Gaussian observation is given by \cite{kay_fundamentals_1998}
\begin{multline} \label{eqn:FI}
    \mathcal{F} = \frac{    \partial \langle \vec{\hat X}^T \rangle}{\partial \theta} 
    \left(     \frac{1}{\mu}
    \mathbf{\Gamma} \right)^{-1}
 \frac{    \partial \langle \vec{\hat X} \rangle}{\partial \theta}
 +\\
 +\frac{1}{2} \operatorname{Tr} \left[ 
   \left(     \frac{1}{\mu}
    \mathbf{\Gamma} \right)^{-1}
     \frac{    \partial  \left(     \frac{1}{\mu}
    \mathbf{\Gamma} \right) }{\partial \theta}  \left(     \frac{1}{\mu}
    \mathbf{\Gamma} \right)^{-1}
     \frac{    \partial   \left(     \frac{1}{\mu}
    \mathbf{\Gamma} \right) }{\partial \theta} 
    \right] =\\
    = \mu M + \frac{1}{2} \operatorname{Tr} \left[  
    \mathbf{\Gamma} ^{-1}
     \frac{    \partial  
    \mathbf{\Gamma}  }{\partial \theta} 
    \mathbf{\Gamma} ^{-1}
     \frac{    \partial  
    \mathbf{\Gamma} }{\partial \theta} 
    \right],
\end{multline}   
where the derivatives of the vectors and matrices are defined elementwise, and the sensitivity $M$ is introduced in Eq.~\eqref{eqn:sensitivity}. For $\mu \gg 1$, the first term in Eq.~\eqref{eqn:FI} dominates the second, which consequently can be neglected. Then, the Cramér-Rao bound \cite{cramer_mathematical_1999} immediately gives
\begin{equation} \label{eqn:CRB}
    \Delta^2 \tilde \theta \ge \mathcal{F}^{-1} = \frac{1}{\mu M},
\end{equation}
proving the inequality \eqref{eqn:boundM}.

Among the estimators in Eq.~\eqref{eqn:generalMoMsetimator}, let us now consider $\tilde \theta $ constructed as a solution of the algebraic equation
\begin{equation} \label{eqn:equation_to_solve}
    \vec c^T ~\vec{\overline{x}}^{(\mu)} = \vec c^T~
\langle \vec{\hat X} \rangle_{\tilde \theta},
\end{equation}
where $\vec c$ is a vector independent on $\tilde \theta$. Thus only the average values $\langle \vec{\hat X} \rangle_{\tilde \theta}$ depend on $\tilde \theta$ in this equation. We assume the function $\langle \vec{\hat X} \rangle_{\tilde \theta}$ can be linearized within the interval $\tilde \theta \pm \Delta \tilde \theta$.
Then the variances of both sides of Eq.~\eqref{eqn:equation_to_solve} are given by
\begin{equation}
    \vec c^T \frac{\mathbf{\Gamma}}{\mu} \vec c = \left(\vec c^T \frac{\partial 
\langle \vec{\hat X} \rangle_{\theta}}{\partial \theta}\right)^2 \Delta^2 \tilde \theta,
\end{equation}
where we used the asymptotic unbiasedness of the estimator $E(\tilde \theta) = \theta$, which follows from the central limit theorem. 
The variance $\Delta^2 \tilde \theta$ of the estimator depends on the choice of the linear coefficients $\vec c$ and takes its minimal value for the coefficients \cite{gessner_metrological_2019}
\begin{equation}
    \vec c =\mathbf{\Gamma}^{-1} 
\frac{\partial \langle \vec{\hat X} \rangle_{\theta}}
{\partial \theta},
\end{equation}
calculated at the true values of the parameter $\theta$. 
In this case, the variance of the estimator is
\begin{align} \label{eqn:var_theta_opt}
   \Delta^2 \tilde \theta = & 
   \frac{1}{\mu} 
   \frac{\vec c^T \mathbf{\Gamma} \vec c}
   {\left(\vec c^T \dfrac{\partial \langle \vec{\hat X} \rangle_{ \theta}}{\partial \theta}\right)^2} =\\=&
    \frac{1}{\mu}
    \frac{
         \dfrac{\partial \langle \vec{\hat X}^T \rangle_{\theta}}{\partial \theta}
         \mathbf{\Gamma}^{-1} 
         \mathbf{\Gamma} 
         \mathbf{\Gamma}^{-1} 
         \dfrac{\partial \langle \vec{\hat X} \rangle_{\theta}}{\partial\theta}}
     {\left(\dfrac{\partial \langle \vec{\hat X}^T \rangle_{\theta}}{\partial \theta} \mathbf{\Gamma}^{-1} \dfrac{\partial \langle \vec{\hat X} \rangle_{ \theta}}{\partial \theta}\right)^2}=\frac{1}{\mu M},
\end{align}
%For a sufficiently large dataset $\mu \gg 1$, the value of the estimator $\tilde{\theta}$ is close to the true value of the parameter $\tilde{\theta} \rightarrow \theta$, and the same is true for the derivatives of the mean values $\partial \langle \vec{\hat{X}} \rangle_{\tilde{\theta}}/\partial \tilde{\theta} \rightarrow \partial \langle \vec{\hat{X}} \rangle_{\theta}/\partial \theta$. In this case, the variance \eqref{eqn:var_theta_opt}
i.e. it saturates the bound \eqref{eqn:CRB}. Thus the estimator $\tilde \theta$ is the optimal one from the class \eqref{eqn:generalMoMsetimator}. Moreover, since the estimator is constructed as a solution of Eq.~\eqref{eqn:equation_to_solve}, it relies solely on the linear combination of the sample means
\begin{equation}
    \tilde \theta = g(\vec c^T ~\vec{\overline{x}}^{(\mu)}).
\end{equation}

\section{Coherency matrix of dipole emission} \label{appendix1}
In the far field of the dipole, the positive-frequency part of the emitted field operator in point $\vec{r}$ reads \cite{scully_quantum_1997}
\begin{equation} \label{eqn:E_dipole}
   \hat{\vec{E}}(\vec{r}, t) \propto \frac{\vec{r} \times [\vec{r} \times \vec{p}]}{|\vec{r}|^3} \hat{\sigma}^{-}(t),
\end{equation}
where $\vec{p}$ is a dipole moment. 

This emission mode does not have the spherical symmetry, however, in the far field, it is locally indistinguishable from a spherical wave. Thus in a paraxial approximation, dipole emitters can be considered as point sources. 

Due to the proportionality \eqref{eqn:E_dipole} between the emitted field operator $\hat{\vec{E}}$ and the dipole transition operators $\hat \sigma^-$, the coherency matrix of the emitted field can be expressed as \eqref{eqn:coh2}, where all the constant factors, except for $\Delta t$, are omitted since they will later be included in the transmissivity factor $\kappa$. This factor represents the coupling between the emission modes and image modes of individual sources. Explicit calculation of \eqref{eqn:coh2}, giving the evolution \eqref{eqn:dipole_dynamic}, result in
\begin{widetext}
\begin{equation} \label{eqn:N_dot_appendix}
    \dot{N}=\langle \hat \sigma^+_1 \hat \sigma^-_1 \rangle=
    \langle \hat \sigma^+_2 \hat \sigma^-_2 \rangle=
   \frac{e^{- g_{11} t}}{2 \left(g_{11}^2-g_{12}^2\right) } 
   \Big(
   \left(g_{11}+g_{12}\right){}^2 e^{-g_{12} t}
   +
   \left(g_{11}-g_{12}\right){}^2 e^{g_{12} t}
   -4 g_{12}^2 e^{-g_{11} t} \Big),
\end{equation}
\begin{equation}
\label{eq:Ng}
   \dot{N} \gamma =\langle \hat \sigma^+_1 \hat \sigma^-_2 \rangle=
    \langle \hat \sigma^+_2 \hat \sigma^-_1 \rangle=
    \frac{e^{- g_{11} t}}{2 \left(g_{11}^2-g_{12}^2\right) } 
   \Big(
   \left(g_{11}+g_{12}\right){}^2 e^{-g_{12} t}
   -
   \left(g_{11}-g_{12}\right){}^2 e^{g_{12} t}
   -4 g_{11} g_{12} e^{-g_{11} t} \Big),
\end{equation}
where $g_{jl}$ is defined in \eqref{eqn:g12}. One can see, that in the non-interacting limit $g_{12} \xrightarrow{} 0$ the emission rate decay exponentially $\dot{N} = e^{-t/\tau}$ and the mutual coherence does not appear $\gamma =0$. Note that the unitary component of the master equation \eqref{eqn:dipole_dynamic}, featuring a collective frequency shift $f_{12}$, does not impact the coherency matrix of the excited modes. Nevertheless, the spectra of the excited modes are influenced by $f_{12}$. Consequently, in instances of frequency-resolving measurement, the dependence $f_{12}(d)$ introduces an additional mechanism for parameter encoding and, in general, enhances sensitivity.
\end{widetext}
\bibliography{referencesBibtex}

%\newpage
%\input{appendix.tex}

  \end{document}